\documentclass[%
 reprint,
 amsmath,amssymb,
 aps,
]{revtex4-2}

\usepackage[T1]{fontenc}
\usepackage[latin9]{inputenc}
\usepackage{amsbsy}
\usepackage{amstext}
\usepackage{amssymb}
\usepackage{graphicx}% Include figure files
\usepackage{dcolumn}% Align table columns on decimal point
\usepackage{bm}
\usepackage{upgreek}
\usepackage{xcolor}
\usepackage{orcidlink}
\hypersetup{pdftitle={Your Title},
	pdfauthor={Your Name},
	pdfpagelayout=OneColumn, pdfnewwindow=true, pdfstartview=XYZ, plainpages=false}

\makeatletter

\begin{document}

\preprint{APS/123-QED}

\title{Array oscillator in coupled waveguides with  nonlinear gain and radiation resistances saturating at exceptional point}% Force line breaks with \\

\author{Alireza~Nikzamir\orcidlink{0000-0003-0111-0220}}
\author{Albert~Herrero-Parareda\orcidlink{0000-0002-8501-5775}}
\author{Nathaniel~Furman\orcidlink{0000-0001-7896-2929}}
\author{Benjamin~Bradshaw\orcidlink{0009-0007-8365-3718}}
\author{Miguel~Saavedra-Melo\orcidlink{0000-0002-8832-9892}}
\author{Filippo~Capolino\orcidlink{0000-0003-0758-6182}}%
 \email{f.capolino@uci.edu}
\affiliation{Department	of Electrical Engineering and Computer Science, University of California, Irvine, CA 92697 USA}

\begin{abstract}

A periodically loaded waveguide composed of periodic discrete nonlinear gain and radiating elements supports a stable oscillation regime related to the presence of an exceptional point of degeneracy~(EPD). After reaching saturation, the EPD in the system establishes the oscillation frequency. We demonstrate a synchronization regime at a stable oscillation frequency, resulting in uniform saturated gain across the array and uniform radiating power. Unlike conventional one-dimensional cavity resonances, the oscillation frequency is independent of the array length. Our investigations further show that when small-signal gain is non-uniformly distributed across the array, the saturated gain results in having a uniform distribution at a gain value that generates an EPD. Experimental validation using the measured board confirmed that the system saturates at an EPD, with a measured spectrum exhibiting very low phase noise. This low noise allows for operation at a clean oscillation frequency. Additionally, the measured uniform power across the array corresponds to the simulation results. The proposed scheme can pave the way for a new generation of high-power radiating arrays with distributed active elements.

\end{abstract}

\maketitle

\begin{figure}[t]
		\begin{centering}
			\includegraphics[width=3.4in]{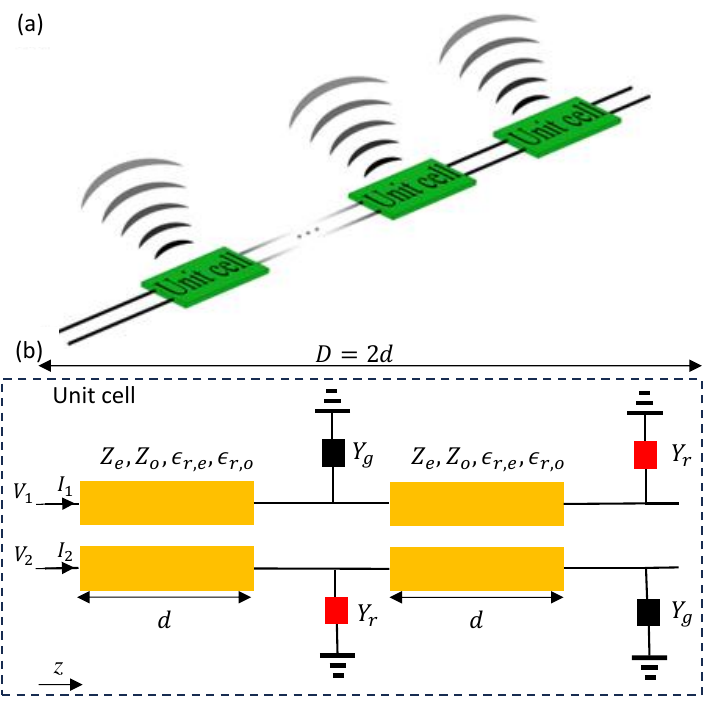}
			\par\end{centering}
		\caption{ (a) Periodic array with elements radiating synchronously while oscillating. Power is radiated by the $Y_r$ elements, two per unit cell, representing antennas. (b) Schematic of a unit cell with length $D=2d$ made of two coupled transmission line~(CTL) segments. The CTLs are characterized by even and odd mode impedances and effective permittivities $Z_{e}$, $Z_{o}$ and $\epsilon_{r,e}$, $\epsilon_{r,o}$, respectively. The CTL is periodically loaded with both a lossy shunt element $Y_{r}$ representing a radiator (e.g., an antenna) and a shunt nonlinear gain element $Y_{g}$ at the same position $z$. The glide symmetry shift length is $d$. Voltages $V_1$ and $V_2$ are given with respect to the ground, not shown for simplicity. \label{fig:Waveguide_linear}}
	\end{figure}

\section{Introduction}

Exceptional points of degeneracy~(EPD) in a system are singular points in a parameter space where some eigenstates experience a "full" degeneracy, i.e., when their eigenvectors and eigenvalues experience degeneracy \cite{Vishik_1960_Solution, Lancaster1964On,Kato1966Perturbation,Heiss1990Avoided,Seyranian1993Sensitivity}. We stress that we deal with an exceptional "degeneracy" as noted in Ref.~\cite{Berry2004Physics}, and that is the reason we have added the "D" in EPD. In the waveguide implementation studied in this paper, the eigenvalues correspond to the modal wavenumbers, and the eigenvectors represent the associated polarization states, here in terms of voltages and currents, that coalesce at an EPD. In general, an EPD occurs in systems that are periodic in space \cite{Figotin2005Gigantic, Abdelshafy2019Exceptional, Yazdi2021Triple}, in time \cite{kazemi2022experimental,rouhi2020exceptional}, in systems that are uniform and have spatial dispersion \cite{Mealy2020General}, or that have gain and loss \cite{Othman2017Theory,Abdelshafy2019Exceptional}. The class of EPD structures that have been mainly studied in the literature is based on parity-time~(PT) symmetry \cite{Bender1998Real,El_Ganainy2007Theory,Guo2009Observation,ruter2010observation,Barashenkov2013PT,Hodaei2014Parity,Schnabel2017PT-symmetric} 
 with balanced gain and loss 
 \cite{heiss_2012physics, ruter2010observation,Guo2009Observation, Kazemi2022HighSensitive, moncada2024frequency}.  
The general conditions for EPD formation in a periodically loaded waveguide with discrete lossy elements and saturable gain were studied in Ref.~\cite{Abdelshafy2021Exceptional}, showing self-standing oscillations. Subsequent studies including nonlinearity in the array through discrete gain elements \cite{Nikzamir2024Exceptional} focused on the saturation regime performance. It was found that the system converges to an EPD while undergoing saturation. However, the saturated gain value would vanish when increasing the length of the array. The purpose of this paper is to show that there are array oscillators where the saturated gain converges to a {\em nonvanishing value} related to an EPD. This demonstration has strong implications for high-power radiation when the length of the array of antennas increases.

\begin{figure*}[t]
		\begin{centering}
			\includegraphics[width=7in]{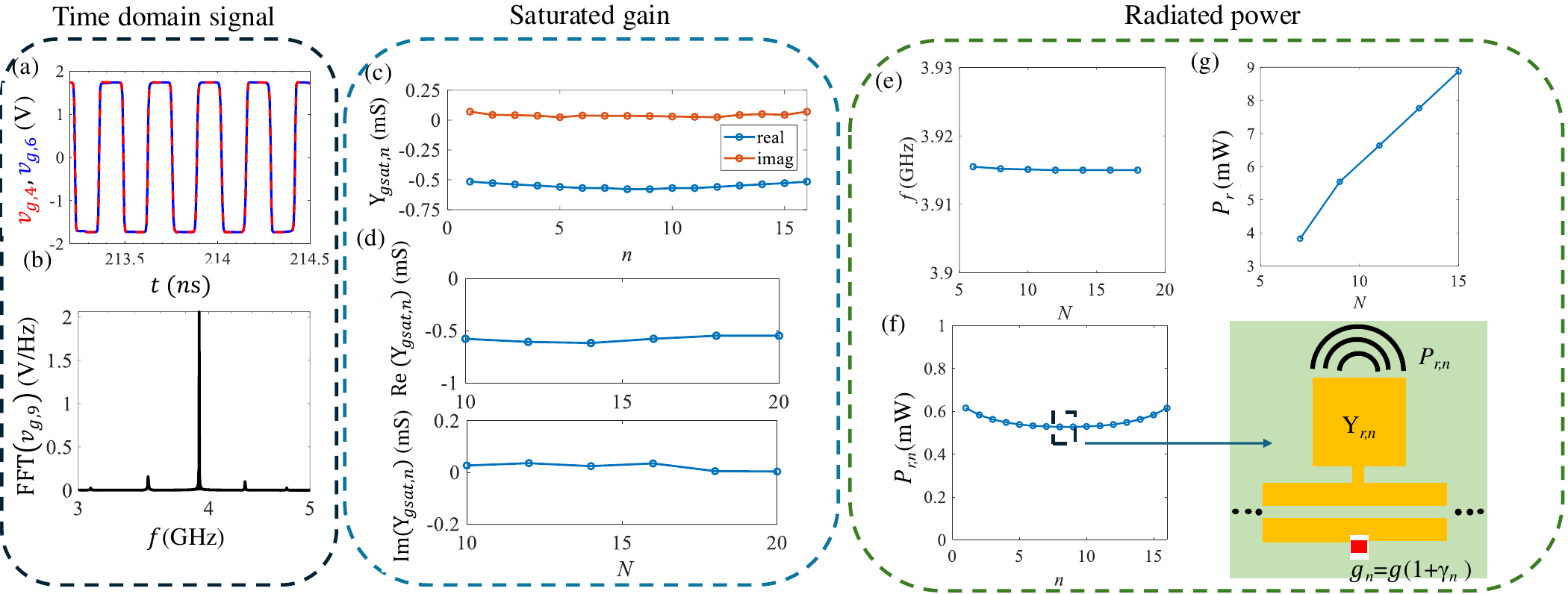}
			\par\end{centering}
		\caption{(a) Time domain signal in the saturation regime for a system with $16$ unit cells in the array showing zero phase shift between voltages of nonlinear gain elements $v_{g,4}$ and $v_{g,6}$ with small signal gain $g=10\:\text{mS}$. (b) The frequency spectrum of voltage $v_{g,9}$ (near the middle position of the system) showing an oscillation frequency of $f_\text{osc}=3.91\:\text{GHz}$ in the saturation regime. (c) Calculated real and imaginary parts of the saturated gain along the structure highlighting a uniform saturated gain around $g_{sat}=0.51\:\text{mS}$ and near negligible imaginary part. (d) The calculated real and imaginary parts of the saturated gain in the middle position of the structure ($n=N/2$) for different structure lengths $N$, showing a trend where the saturated gain approaches $g_{sat}=0.51\:\text{mS}$ and $\text{Im}(Y_{gsat}\approx0)$. (e) Oscillation frequency with respect to structure length, showing a remarkably stable frequency. A total change of $0.01 \%$ in oscillation frequency was observed when increasing the number of array elements from $N=6$ to $N=18$. (g) Total radiated power from the arrays for longer structure lengths shows that the radiated power more than doubles when increasing the array length from $N=6$ to $N=15$. (f) The radiated power of each array element along the structure for $N=16$. There is near uniform power in the middle of the array. In the inset, half of the unit cell is displayed schematically and the radiation power from the $n^{th}$ array is denoted as $P_{r,n}$.
        \label{fig:timedomain}}
	\end{figure*}

Recently, there has been an increasing interest in exploring the effects of nonlinearities in systems that support different kinds of EPD conditions. Studies have examined how nonlinear effects impact PT symmetry in lasing systems \cite{Konotop2016Nonlinear}, EPD-based sensors with saturable gain \cite{nikzamir2022highly, Kananian2020Coupling}, oscillator arrays  \cite{Nikzamir2024Exceptional, Abdelshafy2021Distributed}, oscillator arrays based on degenerate band edge in microstrip circuit \cite{Oshmarin2021experimental}, and noise-resilient systems based on operation near an EPD \cite{suntharalingam2023noise}. Among these EPD-based applications, our focus is on the exceptional degeneracy conditions in waveguides and on what possible advantages they offer. Specifically, we aim to explore the effects of nonlinearities in the saturation regime to design a periodic oscillator based on a waveguide with an array of antennas for efficient high-power radiation with a narrow spectral linewidth. Furthermore, it is important to confirm that such an array system should effectively operate under challenging conditions, such as system tolerances and partial failure. This situation includes the case where all the array elements do not have the same gain or one of the array elements fails or is damaged. To use the proposed concept in real life, the study of robustness and the system's resistance to failed components and perturbations is essential. 

To address these challenges of robust stable oscillation, some of the present authors recently introduced a waveguide with loaded discrete nonlinear gain and antennas (simply represented by linear lossy elements for simplicity) that supports steady oscillation at a second-order EPD \cite{Nikzamir2024Exceptional}. The EPD was the favorable regime of operation after reaching saturation. However, the saturated gain would vanish when the array length was growing to infinity. That finding raised a question that would be important for high-power radiating arrays: is it possible to have an array system supporting EPD with loss and gain where the saturated gain does not decrease with increasing array length?    
 
This paper provides an affirmative answer to such a question that may lead to important applications. We propose a new method of designing arrays that saturate at an EPD with a non-vanishing saturated gain value, resulting in a stable oscillation frequency with increasing radiated power for longer arrays. Remarkably, while undergoing saturation, the gain in the radiating and oscillating array converges to operate at an EPD frequency at steady-state. We also show that this dynamic is independent of the value of the small-signal gain of the discrete active elements.
Additionally, we also investigate the resilience of the oscillation frequency to element failure and variations in the small-signal gain and loss values. This feature is important for practical applications where guaranteeing a constant small-signal gain level across the array is very challenging. Operating at an EPD generally leads to high radiation power and high synchronization of the signal over the radiating elements of the waveguide. It is due to the excitation of an exceptionally degenerate mode, which also leads to a low phase noise performance.

\section{Active Waveguide Array in Saturated Oscillatory Regime}

We analyze the self-oscillation mechanism due to instability in a waveguide made of two coupled microstrips periodically loaded with discrete nonlinear gain and radiating elements (modeled simply as lossy loads), as shown in Fig.~\ref{fig:Waveguide_linear}(b). In each unit cell of length $D$, there are two nonlinear gain elements $Y_g$ and two linear lossy loads $Y_r$, which alternate locations along the waveguide. Hence, the periodic waveguide has glide symmetry as defined in Ref.~\cite{Hessel1973Propagation}, with an effective period of $d=D/2$. Indeed, the structure is equal to itself when a half unit cell shift $d$ is followed by a mirror operation in a transverse plane \cite{Hessel1973Propagation}. The first array case treated in Secs.~\ref{sec:nonlinear} and \ref{sec:EPD}  is neither PT symmetric nor glide-time~(GT) symmetric. The concept of GT symmetry is defined and further explored in Ref.~\cite{Yazdi2021Third}. In Secs.~\ref{sec:nonlinear} and \ref{sec:EPD}, we study the tendency of the steady-state system to operate at an EPD associated to the gain values after reaching saturation. In simulations, the nonlinearity is provided by a cubic $i-v$ curve. In the following, we show that this waveguide ensures stable oscillation frequency at an EPD related to a non-zero saturated gain  $Y_{gsat}$ of the active elements, overcoming the challenges outlined above. Additionally, the ``radiated'' power (i.e., the power dissipated at the lossy loads $Y_r$) is mostly uniform across the array. The waveguide's unit cell is depicted in Fig.~\ref{fig:Waveguide_linear}(b).  In Sec.~\ref{Sec:ExperimentalEPD} we analyze a fabricated prototype to experimentally prove the discovered concepts. The experimented array in Sec.\ref{Sec:ExperimentalEPD} is GT-symmetric in the linear small-signal regime, and we show that it converges to a non GT-symmetric saturated regime. Gain is provided by negative resistances built using operational amplifiers (op amps). For convenience, the experiment in Sec.~\ref{Sec:ExperimentalEPD} is carried out at lower frequencies than the design introduced in Sec.~\ref{sec:nonlinear}, and the coupling between the two microstrips is enhanced by adding a capacitance $Y_c$ in each unit cell of the array. Furthermore, we investigate the stability of the oscillation frequency and system performance when the system is subjected to element failure and under variations in loss and gain values across the array.

\section{Nonlinear simulation results}\label{sec:nonlinear}

\begin{figure*}[t]
		\begin{centering}
			\includegraphics[width=7in]{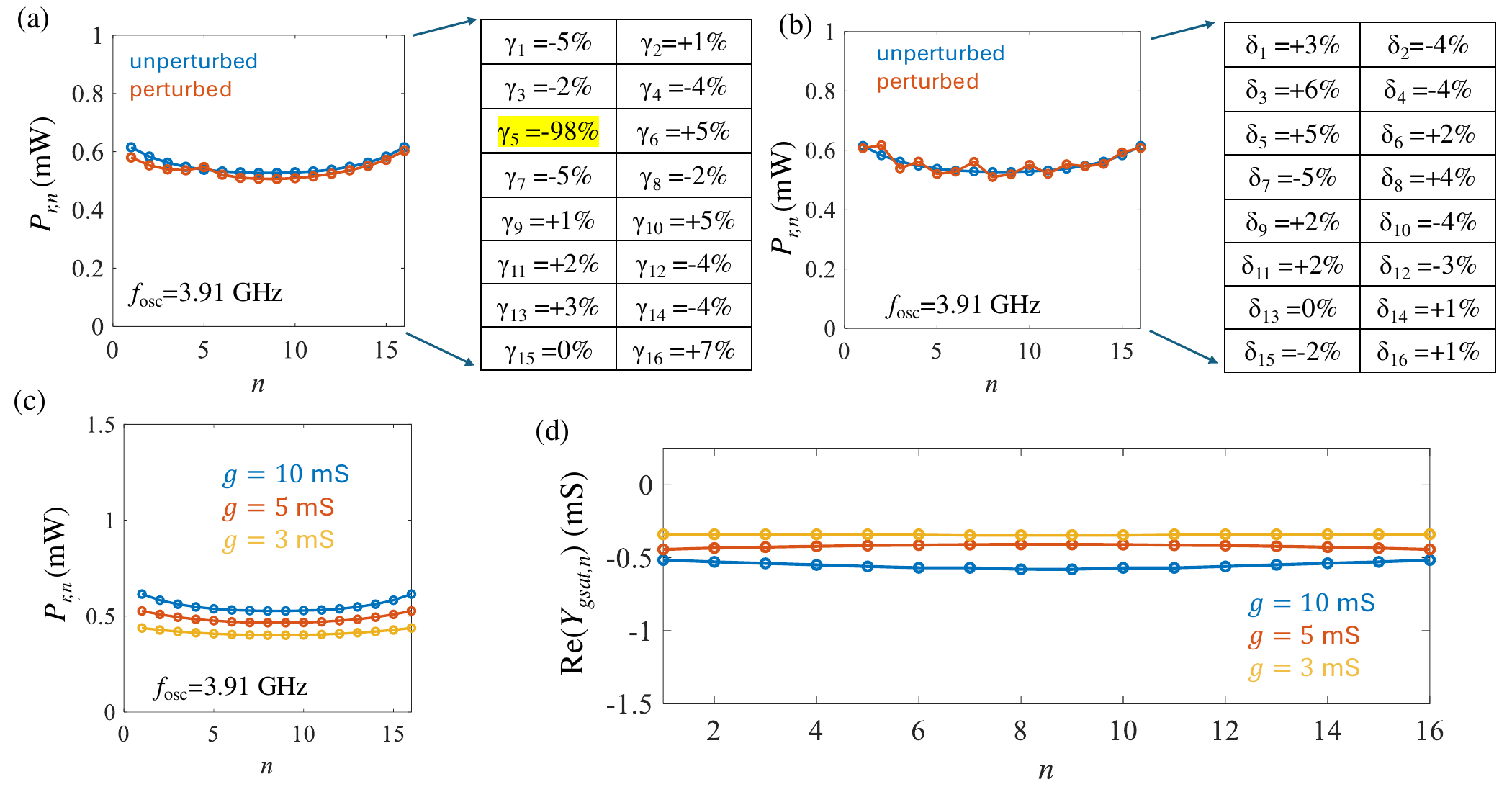}
			\par\end{centering}
		\caption{(a) Radiated power along the structure when the small-signal gain is nonuniform with random values $\gamma_n$ shown as an inset. The $\gamma_5=-98 \%$ value is associated with a faulty nonlinear element. (b) Radiated power along the structure with a random perturbation on the array's admittances $-5\%<\delta_n<5\%$ for the small gain $g=10\:\text{mS}$. Radiated power for the perturbed array's admittances in orange follows the uniform one in blue well. (c) Radiated power for different small-signal gain values along the structure. (d) Saturated gain of each nonlinear gain element for different small-signal gain values. The calculated saturated gains are uniform along the structure, with the calculated value closely approaching the reported saturated gain of $g=0.51\:\text{mS}$ in the middle of the structure. \label{fig:perturbation}}
	\end{figure*}

Nonlinearities play an important role in radio-frequency~(RF) and microwave systems and bring both complexity and benefits \cite{maas2003nonlinear}. In our theoretical analysis, we use time-domain simulations by Keysight Advanced Design System~(ADS) circuit simulator to study the operation of the periodic waveguide in the saturated regime. The nonlinearity of the $n$th element is modeled via the $i-v$ cubic model, as described in Ref.~\cite{Nikzamir2024Exceptional}, through a negative small-signal conductance $Y_{g,n}=-g_n$. The current $i_{n}$ and voltage $v_{n}$ relationship at the discrete nonlinear gain elements in the ADS simulator is defined as

\begin{equation}
    i_n=-g_{n}v_n+\alpha_n v_n^{3},
    \label{eq:Nonlinear_gain}
\end{equation}
\noindent
where $\alpha_n=g_{n}/3$ (unit of ${\rm S}/{\rm V}^2$) describes the saturation level with a $1$ volt turning point. As it will become clear later on, the system is unstable and it oscillates at a frequency $f_{\text{osc}}$ after reaching saturation. With the time-domain results obtained from ADS, we calculate the complex saturated gain admittance in the phasor domain using the Fourier transform of the voltage and current at the oscillation frequency $f_{\text{osc}}$. We implicitly assume the time convention $e^{j\omega t}$. Using the fast Fourier transform~(FFT), the magnitude and phase of the admittance are calculated as 
	
\begin{equation}
     \begin{array}{c}
        \left|Y_{gsat,n}\right|=\frac{\left|\mathrm{FFT}\left(i_{n}\right)\right|_{f_{osc}}}{\left|\mathrm{FFT}\left(v_{n}\right)\right|_{f_\text{osc}}}\\
        \\
        \angle Y_{gsat,n}=\left.\left(\angle\mathrm{FFT}\left(i_{n}\right)-\angle\mathrm{FFT}\left(v_{n}\right)\right)\right|_{f_\text{osc}}
    \end{array},
    \label{eq:Effective_gain}
\end{equation}
\noindent
where $\left|\,\right|$ represents the absolute value, and $\angle$ represents the phase. We performed time-domain simulations for a waveguide consisting of $8$ unit cells (equivalent to $N=16$ nonlinear elements) with a small-signal gain of $g=10\:\text{mS}$ and a loss of $Y_r=40\:\text{mS}$ (representing radiation).
 
In our design, we set the microstrips to have a width $W = 10\:\text{mm}$, a length $d= 198\:\text{mm}$, and spacing between the CTL pair to be $S = 0.3\:\text{mm}$. This period can be designed to have different lengths. It could be larger than a wavelength or even much smaller than a wavelength (see the experimental results later in this paper). It has to be long enough to couple the waves in the two waveguides. The waveguides are implemented on an FR4 substrate with a relative permittivity $\epsilon_r = 4.2$ and a substrate height $h=2\:\text{mm}$. The microstrip is designed using $1$ ounce copper, equivalent to a thickness $t\approx 35\: \mathrm{\upmu m}$. Using the Lincalc tool in Keysight ADS for coupled microstrip lines, we calculated the electrical values of the even and odd modes in the CTL at a frequency of $1\:\text{GHz}$. The relative permittivities for the even and odd modes are $\epsilon_{r,e} = 3.75$, $\epsilon_{r,o} = 3.02$, derived from $\epsilon_{r,e}=(c/v_{p,e})^2$ and $\epsilon_{r,o}=(c/v_{p,o})^2$, where $c$ is the speed of light and $v_p$ is the phase velocity of the respective mode. The corresponding characteristic impedances are $Z_e = 29.8\:\Omega$, and $Z_o = 19.8\:\Omega$. Additional information on the electrical parameters of coupled microstrip lines derived from their physical dimensions is discussed in Ref.~\cite{Hammerstad1980Accurate}.

To perform the finite-length simulation in Keysight ADS, the array is constructed by connecting the unit cells in series. An additional coupled transmission line segment with a length of $d$ is added at the right end of the structure. Consequently, the array begins and ends with coupled transmission lines of length $d$, both terminated with open circuits. After conducting the time domain simulation in Keysight ADS, the voltage signals of the system are shown in Fig.~\ref{fig:timedomain}(a), with the corresponding frequency spectrum shown in Fig.~\ref{fig:timedomain}(b). The transient period is when a system transitions to a steady state, while the rising time indicates how long it takes to reach saturation. The time duration of the transient regime of the system is directly related to the initial small-signal gain. For a value of $g=10\:\text{mS}$, it takes $150\:\text{ns}$ for the system to reach saturation.

The time domain result in Fig.~\ref{fig:timedomain}(a) shows zero phase shift between the nonlinear gain voltages $v_{g,4}$ and $v_{g,6}$ (out of 16 of them) in two consecutive unit cells (of size $D$) in the saturated regime. This means the periodic waveguide operates with zero Floquet-Bloch wavenumber, i.e., $k=0$, at that frequency. The frequency spectrum of the middle nonlinear gain element $v_{g,8}$ in the saturation regime shows an oscillation frequency of $3.91\:\text{GHz}$. The frequency spectrum is calculated by applying the FFT on the saturated signal in the time window from $0.1\:\mu \text{s}$ to $1\:\mu \text{s}$. We used $10^6$ points in a frequency window of $0.1$ GHz to $5$ GHz. 

To better characterize the saturated oscillation regime, in Fig.~\ref{fig:timedomain}(c) we show the real and imaginary parts of saturated gain $Y_{gsat,n}$, where $n=1,2,..., N$, over the $N=16$ active elements using Eq.~(\ref{eq:Effective_gain}). The saturated gain is mostly uniform across all the nonlinear active elements, suggesting a high level of synchronization across the structure. While the system starts with all elements having a small-signal gain of $g=10\:\text{mS}$, it operates with a saturated gain $g_{sat}$ close to $0.51\;\text{mS}$. 

To further investigate this behavior, in Fig.~\ref{fig:timedomain}(d) we examine the saturated gain in the middle array element, $Y_{gsat,n}$, where $n = N/2$, varying the total (even) number of nonlinear elements $N$. There are $N/2$ unit cells, each with two gain elements. For the longer structures, the saturated gain tends to a purely real value, $g_{sat}=0.51\;\text{mS}$ and the saturated reactance vanishes. 

Additionally, this array displays an exceptionally stable oscillation frequency. This feature is depicted in Fig.~\ref{fig:timedomain}(e), which shows the oscillation frequency at the middle of the array for lengths ranging from $N=6$ to $N=18$ nonlinear elements. The oscillation frequency only changes by $0.01 \%$ when the array length is tripled, thus demonstrating its stability against variations in the overall array length.

We analyze the power performance of the array by defining the total time-average ``radiated'' power of the array as the sum of the emitted powers at each radiating element, i.e., $P_{r}=\sum_{n=1}^{N}P_{r,n}$, where $P_{r,n}$ is the time-average power radiated from each array in the steady-state regime. In the saturated regime (i.e., the steady-state regime), we calculate the total time-average power $P_{r}$ radiated by the radiating elements $Y_r$, for growing array length $N$. The results are depicted in Fig.~\ref{fig:timedomain}(g), showing that longer structures radiate more power. Moreover, the power distribution along the length of an array is very important for many applications. Figure~\ref{fig:timedomain}(f) and its inset show the time-average radiated power $P_{r,n}$ by each element $n$ in an array of $N=16$ nonlinear gain elements. The radiating power tends to be uniform in the middle of the array, and it does not vary much over the whole array, in agreement with the observed uniform saturated gain shown in Fig.~\ref{fig:timedomain}(c).

As outlined in Ref.~\cite{Nikzamir2024Exceptional}, the EPD of the infinitely long array (in this limit, the eigenvalue \textit{is} the wavenumber) was the desirable point of operation after saturation. In Ref.~\cite{Nikzamir2024Exceptional}, it was observed that when the system saturates at an EPD, the associated saturated gains of the array elements approached zero when the array length $N$ was increasing. For that array system, the total radiated power grows with array length, but it tends to flatten with growing $N$ because the saturated gain values in the whole array vanish. Notably, we have demonstrated here that it is possible to have systems where the power grows linearly with array length (Fig.~\ref{fig:timedomain}(g)), hence with saturated gain values that tend to a constant for growing array length $N$. And, importantly, the saturated regime is still associated to an EPD of the infinitely long array. This leads to a stable frequency of oscillation and low phase noise, as demonstrated next.
 
\subsection{Gain and Loss Perturbation and Failure Analysis}\label{sec:Perturbation}

RF or microwave oscillator schemes are required to have a stable oscillation frequency \cite{Frerking1978Crystal,Walls1986Measurements}, a high-quality factor \cite{An_Sun_Hyun1999K-band,Hosoya2000lowphase-noise}, a strong independence with respect to the loads \cite{Abdelshafy2021Distributed}, and a high output power \cite{Kasagi2019Large-scale}. To confirm these dynamic properties and to confirm the robustness of our proposed structure with respect to fabrication errors and tolerances, we examine a scenario where the small-signal gain $g$ in each unit cell varies randomly within a certain range. The reason for this study is due to the greater difficulty in designing active elements than the passive ones. Therefore, we initially focus on perturbations of the values of the active elements, i.e., the nonlinear gain. Specifically, the gain varies as $g_{n}= 10\times(1+\gamma_{n})\;\text{mS}$, with $n=1,2,..., N$ and where the $-10\%<\gamma_{n}<10\%$ values are shown in Fig.~\ref{fig:perturbation}(a). The $\gamma_n$ values follow a uniform random distribution, as assumed in Ref.~\cite{Nikzamir2024Exceptional}. 

We also study the resilience of the system to a damaged active element by setting $\gamma_{5}=-98\%$, which is equivalent to an element with near-zero small-signal gain. The resulting power delivered to each radiating element $Y_r$, depicted in Fig.~\ref{fig:perturbation}(a), shows that even with tolerances in the small-signal gain $g_n$ and with faulty elements, the array operates at the desired oscillation frequency $f_\text{osc}=3.91\;\text{GHz}$, and power delivered to all the loads $Y_r$ follows a similar pattern as the unperturbed case where $\gamma_n=0$. Also, the power radiated by $Y_{r,5}$, the lossy element adjacent to the faulty gain element, does not change.

Since lossy elements generally display lower tolerances than active elements, we assume that the arrays' admittances have a $5\%$ variation around a nominal value $Y_r=40\:\text{mS}$, i.e., they are $Y_{r,n}= (1+\delta_{n}) \times 40\:\text{mS}$. Even with the values displayed in Fig.~\ref{fig:perturbation}(b) for the perturbed array elements, we observe roughly uniform power radiation along the system.

Additionally, we studied changes in structure performance when the small-signal gain $g$ is chosen to be constant over the entire array, but assuming three different values. Figure~\ref{fig:perturbation}(c) shows the power delivered to the array elements $Y_r$, and Fig.~\ref{fig:perturbation}(d) shows the saturated gain values over the discrete array elements. These results were obtained by performing time-domain simulations with three different values of the small-signal gain: $g=10\:\text{mS}$, $g=5\:\text{mS}$, and $g=3\:\text{mS}$ for all the elements across the array. In all cases, the system oscillates and retains the same stable oscillation frequency of $f_\text{osc}=3.91\;\text{GHz}$ regardless of the chosen small-signal gain value, and it displays roughly the same power distribution across the array. Moreover, the saturated gain values shown in Fig.~\ref{fig:perturbation}(d) are approximately the same as the $g=0.51\:\text{mS}$ in the original design. These results demonstrate that systems with different initial small-signal gain values still saturate close to the same gain value as the original design. 

The saturation behavior of the system is intricately connected to the behavior of its modes. The results described in Sec.~\ref{sec:nonlinear} suggest the saturated system operates at or near an EPD, as previously shown in Ref.~\cite{Nikzamir2024Exceptional}. Oscillators operating at an EPD show a very stable oscillation frequency, that is roughly independent of waveguide length \cite{Abdelshafy2021Distributed}. Additionally, EPD-based oscillators display synchronized radiation across the array, caused by operation at the center or edge of the first Brillouin zone of the Bloch mode dispersion diagram, i.e., with $k = 0$ or $k=\pi/D$, as shown in the next section. We investigate the relationship between the system's modes, their degeneracy, and the overall performance. 

\section{Coupled transmission lines with discrete gain and radiation admittances}\label{sec:EPD}

Exploiting the concept of EPDs to improve the performance of oscillators with distributed gain and loss has been explored in the past (see Refs.~\cite{Abdelshafy2021Exceptional,Mealy2020EPDinEbeam,Mealy2021XbandBWO} for details). Here, as in Ref.~\cite{Abdelshafy2021Exceptional}, the array's radiating elements are simply represented by periodic radiation admittances $Y_r$ along the TL. Though gain elements are nonlinear, in this section we assume they are linear and have their saturated gain value. We analyze the modes supported by the unit cell discussed in Fig.~\ref{fig:Waveguide_linear}(b). The transfer matrix of the CTLs of given length $d$ requires introducing even and odd modes, with propagation constants $\beta_e=2\pi f\sqrt{\epsilon_{r,e}}/c$ and $\beta_o=2\pi f\sqrt{\epsilon_{r,o}}/c$, where $c$ is the speed of light, and characteristic impedances $Z_e$ and $Z_o$. Their mode propagation is synonymously described in Ref.~\cite{Pozar2011Microwave}, but using the forward transfer matrix formalism, we have

\begin{equation}
    \begin{pmatrix}
        V(d) \\
        I(d)
    \end{pmatrix}
    =
    \begin{pmatrix}
        \cos(\beta d) & -jZ \sin(\beta d) \\
        -j \sin(\beta d)/Z & \cos(\beta d)
    \end{pmatrix}
    \begin{pmatrix}
        V(0) \\
        I(0)
    \end{pmatrix},
    \label{eq:TMatrixEvenModes}
\end{equation}
\noindent
where $\beta=\beta_{e,o}$, and $Z=Z_{e,o}$. Furthermore, $V=V_{e,o}$ and $I=I_{e,o}$ are the voltage and currents of the even and odd modes, defined by
\begin{equation}
    \begin{matrix}
        V_e=\frac{V_1+V_2}{2}, I_e=\frac{I_1+I_2}{2} \\
        V_o=\frac{V_1-V_2}{2}, I_o=\frac{I_1-I_2}{2}
    \end{matrix}.
\end{equation}

%Here, we define the state vector of the periodic system as $\begin{pmatrix}V_e, I_e, V_o, I_o\end{pmatrix}^\text{T}$, where the superscript $T$ denotes the transpose operation. This state vector includes the elements for the even and odd modes, which propagate with their respective propagation constants $\beta_e, \beta_o$ and characteristic impedances $Z_e, Z_o$. 

It is convenient to define the waveguide state vector as $\Psi(z) = \begin{pmatrix}V_1, I_1, V_2, I_2\end{pmatrix}^\text{T}$, where the superscript T denotes transpose operation, and the parameters are defined in Fig.~\ref{fig:Waveguide_linear}(b). This formulation is analogous to the one used for electric fields in optical waveguides \cite{nadaDesignModifiedCoupled2023}. The ``evolution'' of the state vector across the CTL segment of length $d$ is given by

\begin{equation}
    \begin{pmatrix}
        V_1(d) \\
        I_1(d) \\
        V_2(d) \\
        I_2(d)
    \end{pmatrix}
    =
    \begin{pmatrix}
        T_{11} & T_{12} & T_{13} & T_{14} \\
        T_{21} & T_{11} & T_{23} & T_{13} \\
        T_{13} & T_{14} & T_{11} & T_{12} \\
        T_{23} & T_{13} & T_{21} & T_{11} \\
    \end{pmatrix}
    \begin{pmatrix}
        V_1(0) \\
        I_1(0) \\
        V_2(0) \\
        I_2(0)
    \end{pmatrix},
\end{equation}
where the $4$x$4$ matrix represents the transfer matrix $\mathbf{T_{\text{CTL}}}$ of the CTL segment. Its elements are

\begin{figure}[t]
		\begin{centering}
			\includegraphics[width=3.4in]{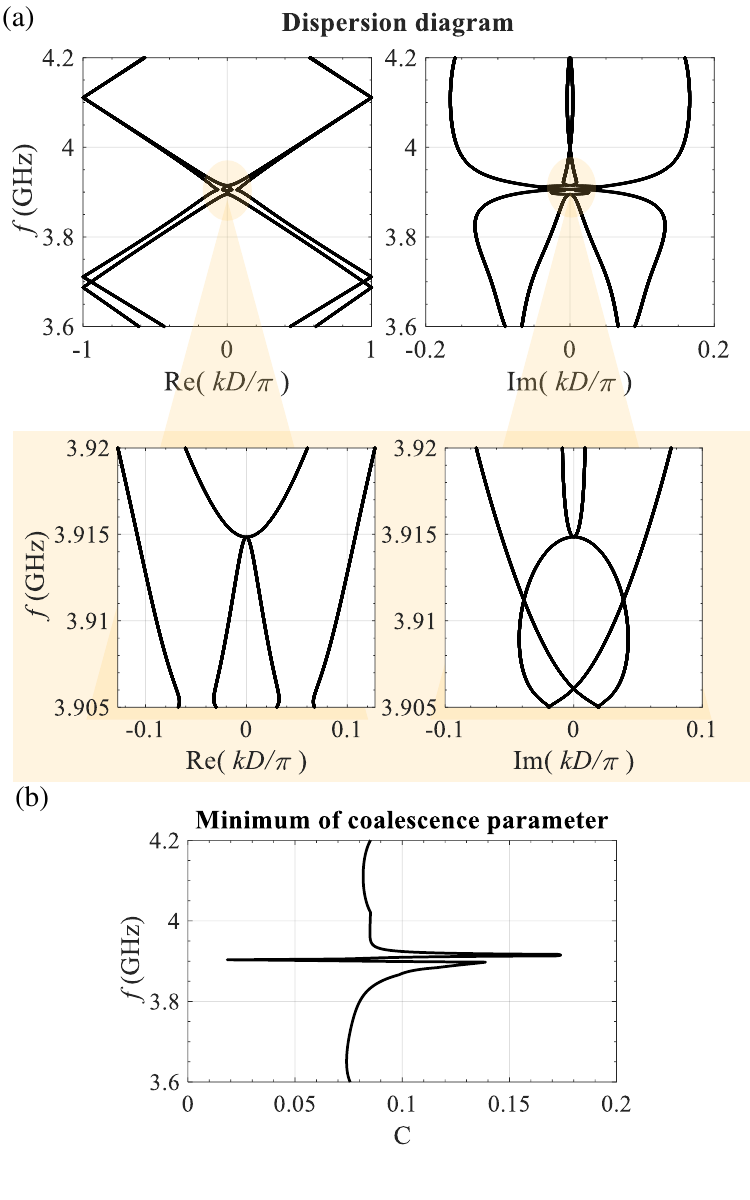}
			\par\end{centering}
		\caption{ (a) Real and imaginary parts of the complex wavenumber $k$ of the modes supported by the infinitely long array versus frequency with a zoomed-in version around the degeneracy at $kD/\pi=0$. Branches coalesce at a second-order EPD at $3.915\:\text{GHz}$. This is the same steady-state oscillation frequency of the array of finite length in the steady state regime (i.e., after saturation), obtained from time domain simulations. (b) The vanishing of the coalescence parameter confirms the coalescence of two eigenvectors at $3.915\:\text{GHz}$. 
  \label{fig:dispersion}}
\end{figure}

\begin{table}[h]
\begin{tabular}{ccl}
    $T_{11}$ & = & $(\cos\theta_e+\cos\theta_o)/2$ \\
    $T_{12}$ & = & $-j(Z_{e} \sin\theta_e + Z_{o} \sin\theta_o)/2$ \\
    $T_{13}$ & = & $(\cos\theta_e-\cos\theta_o)/2$ \\
    $T_{14}$ & = & $-j(Z_{e} \sin\theta_e - Z_{o} \sin\theta_o)/2$ \\
    $T_{21}$ & = & $-j(\sin(\theta_e)/Z_{e} + \sin(\theta_o)/Z_{o} )/2$ \\
    $T_{23}$ & = & $-j(\sin(\theta_e)/Z_{e} - \sin(\theta_o)/Z_{o} )/2$,
\end{tabular}
\end{table}
\noindent
where $\theta_e=\beta_e d$ and $\theta_o=\beta_o d$ are the electric lengths of the even and odd modes of the two uniform waveguide segments, respectively.
The transfer matrix of the unit cell, depicted in Fig.~\ref{fig:Waveguide_linear}(b), is defined as 

\begin{equation}
    \mathbf{T_{\text{U}}}=\left(\begin{array}{cc}
    \mathbf{T_{\text{loss}}} & \mathbf{0}\\
    \mathbf{0} & \mathbf{T_{\text{gain}}}
    \end{array}\right) \mathbf{T_{\text{CTL}}}\left(\begin{array}{cc}
    \mathbf{T_{\text{gain}}} & \mathbf{0}\\
    \mathbf{0} & \mathbf{T_{\text{loss}}}
    \end{array}\right) \mathbf{T_{\text{CTL}}},
\end{equation}
\noindent
where $\mathbf{0}$ is a $2\times2$ matrix of zeros and $\mathbf{T_{\text{gain}}}$, $\mathbf{T_{\text{loss}}}$ are the transfer matrices of the shunt gain and loss elements, i.e., 
\begin{equation}
    \mathbf{T}_{\text{gain}}=\left(\begin{array}{cc}
    1 & 0\\
    -Y_g & 1
    \end{array}\right),\:\:\\
    \mathbf{T}_{\text{loss}}=\left(\begin{array}{cc}
    1 & 0\\
    -Y_r & 1
    \end{array}\right).
\end{equation}

In the small signal regime, we would have $Y_g=-g$. However, for us it is more important to analyze the state of the system after reaching saturation, hence we assume $Y_g=Y_{gsat}$. The analysis in the previous section tells us that $Y_{gsat}=-g_{sat}$ and we can neglect the reactance since it is almost zero.   

The evolution equation for the state vector across a unit cell is given by $\Psi(z+D)=\mathbf{T}_\text{U}\Psi(z)$. The modes supported by the periodic guiding structure are found by applying the Floquet-Bloch theorem \cite{Floquet-Bloch2013Gomez}, resulting in $\mathbf{\Psi}(z+D)=e^{-jkD}\mathbf{\Psi}(z)$ where $D=2d$ is the unit-cell period and $k$ is the Floquet-Bloch wavenumber. Note that $d$ is the length of the CTL segments. The eigenmodes are found by solving the eigenvalue problem, i.e.,

\begin{equation}
    \left(\mathbf{T}_\text{U}-\zeta \mathbf{\underline{\mathrm{I}}}\right)\mathbf{\Psi}=0,
\end{equation}
where $\underline{\mathbf{I}}$ is the identity matrix of order four, $\zeta=e^{-jkD}$ is the eigenvalue of a mode, and $\boldsymbol{\Psi}$ is its associated eigenvector. The four eigenvalues are found by solving the characteristic equation $\mathrm{det}\left(\mathbf{T}_\text{U}-\zeta\mathbf{\underline{\mathrm{I}}}\right)=0$. After some algebra, we obtain

\begin{equation}
    \begin{array}{c}
        \zeta^4 + b\zeta^3 + c\zeta^2 + b\zeta+ 1 = 0,
    \end{array}
    \label{eq:FullDispersionRelation}
\end{equation}
where the coefficients $b$ and $c$ are given in Appendix~A. Because of reciprocity, discussed in Appendix~A, the characteristic equation is equivalent to

\begin{equation}
    (\zeta^{2}-a_{1}\zeta+1)(\zeta^{2}-a_{2}\zeta+1)=0,
    \label{eq:genral_polynomial}
\end{equation}
where $a_1 = \zeta_{1}+1/\zeta_{1}, \: a_2 = \zeta_{2}+1/\zeta_{2}$. The terms $\zeta_{1}$ and $\zeta_{2}$ are the eigenmode solutions for Eq.~(\ref{eq:FullDispersionRelation}), as well as $1/\zeta_{1}$ and $1/\zeta_{2}$ leading to the four wavenumbers $k_1$ and $-k_1$, and $k_2$ and $-k_2$.

In the remainder of this section, we discuss the EPD condition based on the characteristic equation with the coefficients shown in Appendix A. The relationship between the coefficients in the full dispersion equation in Eq.~(\ref{eq:FullDispersionRelation}) and in the general polynomial in Eq.~(\ref{eq:genral_polynomial}) are

\begin{equation}
    \begin{array}{c}
    b=-a_1-a_2\\
    c=2+a_1a_2
    \end{array}.
    \label{eq:bandc}
\end{equation}

The four modes that satisfy Eq.~(\ref{eq:genral_polynomial}) are

\begin{equation}
    \begin{array}{c}
    \zeta_{1,3}=\frac{a_{1}\pm\sqrt{a_{1}^{2}-4}}{2},\:\:\:\:\zeta_{2,4}=\frac{a_{2}\pm\sqrt{a_{2}^{2}-4}}{2}
    \end{array}.
    \label{eq:genreal_modes}
\end{equation}

Based on the above Eq.~(\ref{eq:genreal_modes}), this structure can support exceptional degeneracies (i.e., EPDs) of orders $2$ and $4$ (involving either $2$ or $4$ coalescing modes, respectively). An EPD of order $2$ or $4$ is formed when the modes of the system fall into one of the following cases: (I) when either $\zeta_1=\zeta_3$ or $\zeta_2=\zeta_4$, that is, when either $a_1=\pm2$ or $a_2=\pm2$, and $a_1\neq a_2$. In this case, there is one EPD of order $2$, represented as the point in the dispersion diagram where the slope vanishes, at the center or edges of the Brillouin Zone~(BZ), i.e., $k=0$ when either $a_1$ or $a_2$ is 2, or $k=\pi/D$ when either $a_1$ or $a_2$ is $-2$. The remaining two modes do not merge. (II) When $\zeta_1 = \zeta_3$ and $\zeta_2 = \zeta_4$, yet $\zeta_1 \neq \zeta_2$. This is equivalent to the condition that $a_1 =\pm2$, $a_2=\pm2$, with $a_1 \neq a_2$. There are then two EPDs of order $2$ at the same frequency, where one occurs at the center of the BZ and the other one at its edges. In this case, the dispersion diagram denotes two distinct EPDs at the same frequency, one at $k=0$ and at $k=\pm \pi/D$. (III) When $\zeta_1=\zeta_2$ and $\zeta_3=\zeta_4$, namely, when $a_1=a_2\neq\pm2$. In this instance, there are two EPDs of order $2$ which occur at reciprocal positions off the center or edges of the BZ, i.e., $k_1=k_2$, $k_3 = k_4$, and $k_n \ne 0,\pm\pi/D$. (IV) When $\zeta_1 = \zeta_2$ and $a_1= a_2=\pm2$, the four eigenvalues become degenerate, potentially leading to the formation of a fourth-order EPD at the center or edges of the BZ. In this work, we focus on the formation of EPDs of order $2$ belonging to case I, so cases II, III and IV are not considered in the following.  

Figure~\ref{fig:dispersion}(a) displays the dispersion diagram of the periodic waveguide with the same parameter values in the saturation regime as in Sec.~\ref{sec:nonlinear}, namely, $g_{sat}=0.51\:\text{mS}$, $Y_r=21.3\:\text{mS}$, $d=198\:\text{mm}$, $\epsilon_{r,e}=3.75$, $\epsilon_{r,o}=3.02$, $Z_e=29.8\:\Omega$ and $Z_o=19.8\:\Omega$. As expected from the results in Ref.~\cite{Nikzamir2024Exceptional}, the saturated periodic system displays a second-order EPD, and the frequency at which it occurs corresponds to the frequency at which the finite-length array discussed in the previous section operates after reaching saturation, namely $f_\text{EPD}=f_\text{osc} = 3.915\:\text{GHz}$. At this frequency, we calculate $a_1=2$ and $a_2=1.912+0.228i$, which show that the EPD is at $k=0$.

The occurrence of the exceptional degeneracy is confirmed by using the concept of coalescence parameter $\text{C}$, introduced in Ref.~\cite{Abdelshafy2019Exceptional} where it is referred to as hyperdistance. This parameter $\text{C}$ is a figure of merit to assess how close the infinitely-long array is to an EPD by observing the degree of coalescence of the system's eigenvectors. The coalescence parameter of two eigenvectors $m$ and $n$ is defined as 

\begin{equation}\mathrm{C}_{mn}=|\mathrm{sin}\theta_{mn}|,\:\cos\theta_{mn}=\frac{\left|\left\langle \mathbf{\mathbf{\Psi}}_{m},\mathbf{\mathbf{\Psi}}_{n}\right\rangle \right|}{\left\Vert \mathbf{\mathbf{\Psi}}_{m}\right\Vert \left\Vert \mathbf{\mathbf{\Psi}}_{n}\right\Vert },
\label{eq:coalescence}
\end{equation}
\noindent
where $\theta_{mn}$, with any choice of $n=1,...,4$ and $m=1,...,4$, with $m\ne n$, represents the angle between $2$ eigenvectors $m$ and $n$ in a four-dimensional complex vector space via the inner product $\left\langle \mathbf{\Psi}_{m},\mathbf{\Psi}_{n}\right\rangle =\mathbf{\Psi}_{m}^{\dagger}\mathbf{\Psi}_{n}$, where the dagger symbol $\dagger$ denotes the complex conjugate transpose operation, and $\left\Vert \mathbf{\Psi}\right\Vert =\sqrt{\left\langle \mathbf{\Psi},\mathbf{\Psi}\right\rangle}$ represents the norm of a complex vector \cite{Scharnhorst2001Angles,Nada2021FrozenMode}.

The coalescence parameter defined in Eq.~(\ref{eq:coalescence}) is always positive and smaller than one with small values indicating that the eigenvectors of the structure are close to degeneracy. At an EPD, the transfer matrix of the unit cell $\mathbf{T_{U}}$ is similar to a Jordan Block matrix \cite{hogben2006handbook,Figotin2005Gigantic}. Since the transfer matrix is a $4$x$4$ matrix, it supports four eigenmodes and eigenvectors. We are searching for a minimum coalescence parameter between each pair of eigenvectors $\text{C}=\text{min}_{m,n}\text{(C}_{mn})$. The coalescence parameter is equal to zero when two eigenvectors in the waveguide coalesce, i.e., when the CTL system experiences a second-order EPD. In physical systems and numerical studies, the coalescence will never be exactly zero. Figure~\ref{fig:dispersion}(b) shows the minimum of the coalescence parameter versus frequency. A degeneracy is clearly identified at $f=3.915\:\text{GHz}$, which is the same oscillation frequency associated with the steady-state regime of oscillation with saturated nonlinear gain values shown in Fig.~\ref{fig:timedomain}(b). In conclusion, we have demonstrated via simulations that the finite-length periodic array with nonlinear gain elements saturates and oscillates at an EPD frequency of the waveguide dispersion. The EPD condition can be thought of as a ``destination'' the nonlinear system approached upon saturation.

\section{Experimental confirmation of EPD-based stable oscillation}
\label{Sec:ExperimentalEPD}

To provide an experimental confirmation of a periodic array system that oscillates at an EPD related to saturated nonlinear gain, as shown in the previous sections, we create another design at a lower frequency in the MHz range, instead of the $f_{\text{osc}}=3.91\:\text{GHz}$ previously considered. By operating at lower frequencies, we can neglect several parasitic capacitances and small lengths, and therefore we can test the array using standard wiring. 

The new design shows that the array oscillates at $f_{\text{osc}}=21.67\:\text{MHz}$. To build an array that achieves uniform power distribution across the structure, we consider the same configuration discussed in Fig.~\ref{fig:Waveguide_linear}. Due to the long period of the original unit cell, here we reduce the period $D$ in experimental verification because of fabrication limitations. When decreasing the unit-cell period while maintaining differences between the even and odd modes, a strong coupling between CTLs is needed. Since it is difficult to operate with a very large difference in even and odd mode permittivities $\epsilon_{r,e}$ and $\epsilon_{r,o}$, respectively, here we include additional capacitances between the CTLs next to the lossy and active loads as shown in Fig.~\ref{fig:Measurment_sim}(a). After this modification, we first numerically verify the system's performance, followed by experimental results that validate our findings in Sec.~\ref{sec:nonlinear} and Sec.~\ref{sec:EPD}.

\subsection{Theory and simulation of experimental board}\label{subsec:Simulation_meas}

\begin{figure}[t]
		\begin{centering}
			\includegraphics[width=3.4in]{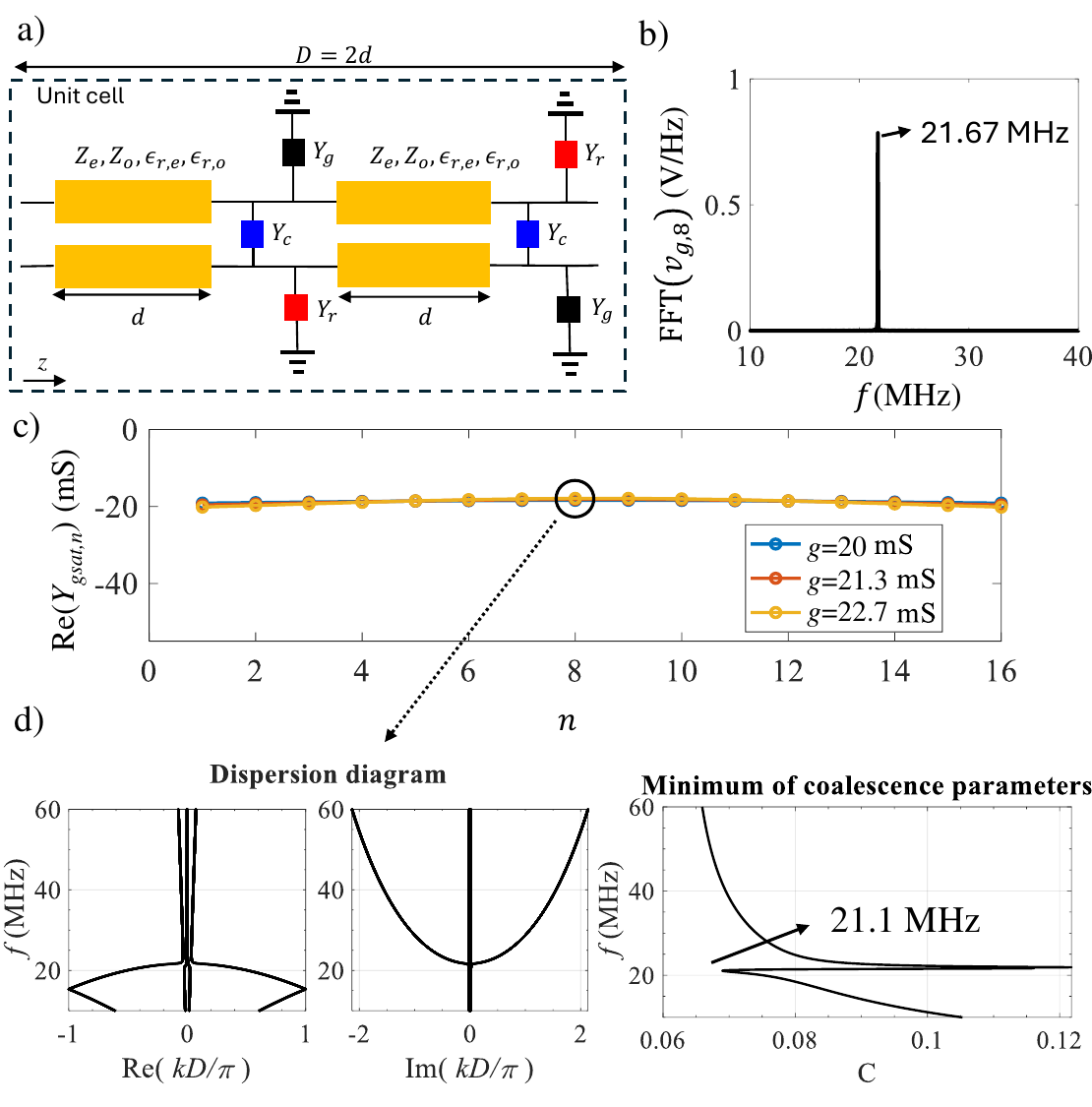}
			\par\end{centering}
		\caption{ (a) Proposed circuit with a coupling capacitive admittance $Y_c = j\omega C$ between the two TLs. (b) Frequency spectrum of the voltage $v_{g,8}$ (near the middle position of the system) after reaching the saturation regime. (c) Saturated gain calculated in the saturation regime over the structure for $N=16$. (d) Calculated dispersion diagram and coalescence parameter for the structure using $g_{sat} = 18.1\:\text{mS}$ as gain, showing an EPD close to the spectrum maximum in (b).
\label{fig:Measurment_sim}}
\end{figure}

The theoretical investigation slightly differs from that in the previous sections because of the two extra capacitances per unit cell shown in Fig.~\ref{fig:Measurment_sim}. We will first show the time-domain simulation results by modeling the nonlinear gain as described earlier in Eq.~(\ref{eq:Nonlinear_gain}) using ADS Keysight. Then, the modal dispersion and EPD are verified using the transfer matrix accounting for the coupling capacitors, given by

\begin{equation}
  \mathbf{T_{\text{U}}}=\left(\begin{array}{cc}
\mathbf{T_{\text{loss}}} & \begin{array}{cc}
0 & 0\\
Y_{c} & 0
\end{array}\\
\begin{array}{cc}
0 & 0\\
Y_{c} & 0
\end{array} & \mathbf{T_{\text{gain}}}
\end{array}\right)\mathbf{T_{\text{CTL}}}\left(\begin{array}{cc}
\mathbf{T_{\text{gain}}} & \begin{array}{cc}
0 & 0\\
Y_{c} & 0
\end{array}\\
\begin{array}{cc}
0 & 0\\
Y_{c} & 0
\end{array} & \mathbf{T_{\text{loss}}}
\end{array}\right)\mathbf{T_{\text{CTL}}},
\end{equation}
\noindent
where $\mathbf{T_{\text{gain}}}$ and $\mathbf{T_{\text{loss}}}$ are the transfer matrices of the shunt gain and loss with the added capacitor, i.e., 

\begin{equation}
\mathbf{T}_{\text{gain}}=\left(\begin{array}{cc}
1 & 0\\
-(Y_g+Y_c) & 1
\end{array}\right),\:\:\\
\mathbf{T}_{\text{loss}}=\left(\begin{array}{cc}
1 & 0\\
-(Y_r+Y_c) & 1
\end{array}\right), 
\end{equation}
\noindent
where $Y_c=j \omega C$ is the coupling capacitive admittance. We use the Rogers RO4350B grounded dielectric substrate with relative permittivity $\epsilon_r = 3.48$ and height $h=2\:\text{mm}$, covered by $1$ ounce copper. It has low loss, characterized by $\text{tan}\delta=0.003$ for frequencies ranging from $0.1\:\text{MHz}$ to $1\:\text{GHz}$. We design the microstrips with a width $W = 10\:\text{mm}$ and a length $d= 50\:\text{mm}$. The spacing between the CTL segments is $S = 1\:\text{mm}$. The unit-cell period is $D=2d = 100\:\text{mm}$, roughly a quarter of the value in the original design. The coupled microstrips are characterized by $\epsilon_{r,e} = 3.2$, $\epsilon_{r,o} = 2.7$, $Z_e = 32\:\Omega$, and $Z_o = 24.4\:\Omega$, at a frequency of $1\:\text{GHz}$.  The loads representing the radiation admittance are  $Y_r=21.3\:\text{mS}$.

To conduct the finite-length simulation in Keysight ADS and ensure consistency with the experimental board being tested later on, the array starts with a coupled transmission line of length $d/2$ in the first unit cell instead of $d$. Similarly, an additional CTL of length $d/2$ is included at the other end. As a result, the system begins and ends with coupled transmission lines of length $d/2$, both terminated with open circuits. Via ADS Keysight simulations, we found that increasing the coupling capacitance from $C=1\:\text{nF}$ to $16\:\text{nF}$, with a small-signal gain of $g=21.3\:\text{mS}$, leads to a significant reduction in the oscillation frequency. To make fabrication and measurement feasible, we chose a coupling capacitance of $C=16\:\text{nF}$, which significantly lowers the oscillation frequency to  $f_\text{osc}=21.67\:\text{MHz}$ for $N=16$ nonlinear gain elements as seen as the first peak in Fig.~\ref{fig:Measurment_sim}(b) that shows the frequency spectrum of the system. Based on the theoretical investigation and time-domain simulation results in Sec.~\ref{sec:nonlinear}, we expect uniformly saturated gain values across the array.

Figure~\ref{fig:Measurment_sim}(c) shows the real part of the saturated gain along the array for three different small-signal gains: $g=20\:\text{mS}$, $g=21.3\:\text{mS}$, and $g=22.7\:\text{mS}$. These results show the expected behavior for the oscillation frequency at $f_\text{osc}=21.67\:\text{MHz}$, with a saturated gain value close to $g_{sat,8}=18.1\:\text{mS}$, in all cases, {\em independently of the value of the small-signal gain} considered. The array primarily oscillates at the fundamental frequency, with the third harmonic appearing $20.4$ dB below the peak at the fundamental frequency $f_\text{osc}=21.67\:\text{MHz}$ in the obtained simulated frequency spectrum for Keysight ADS. Using the calculated value of the eighth element's saturated gain, $g_{sat,8}=18.1\:\text{mS}$, in Fig.~\ref{fig:Measurment_sim}(d), we plot the wavenumber dispersion of the modes in the infinitely long array, and the coalescence parameter \text{C} whose minimum reveals how close the system is to the ideal EPD. It is important to note that this coalescence is determined by finding the minimum among six possible combinations of the four eigenvectors, as outlined in Sec.~\ref{sec:EPD}. The system exhibits a second-order EPD at $21.1\:\text{MHz}$ based on the calculated coalescence parameter, which is close to the frequency of oscillation of the finite-length array with nonlinear elements. In these results, the coalescence parameter does not completely vanish, but it is very close to zero, revealing that the system approaches the ideal EPD of the infinitely long array. By analyzing the dispersion of this updated design, we see that it is a second-order EPD of case (I) as in the previous design.

\begin{figure*}[t]
		\begin{centering}
			\includegraphics[width=7in]{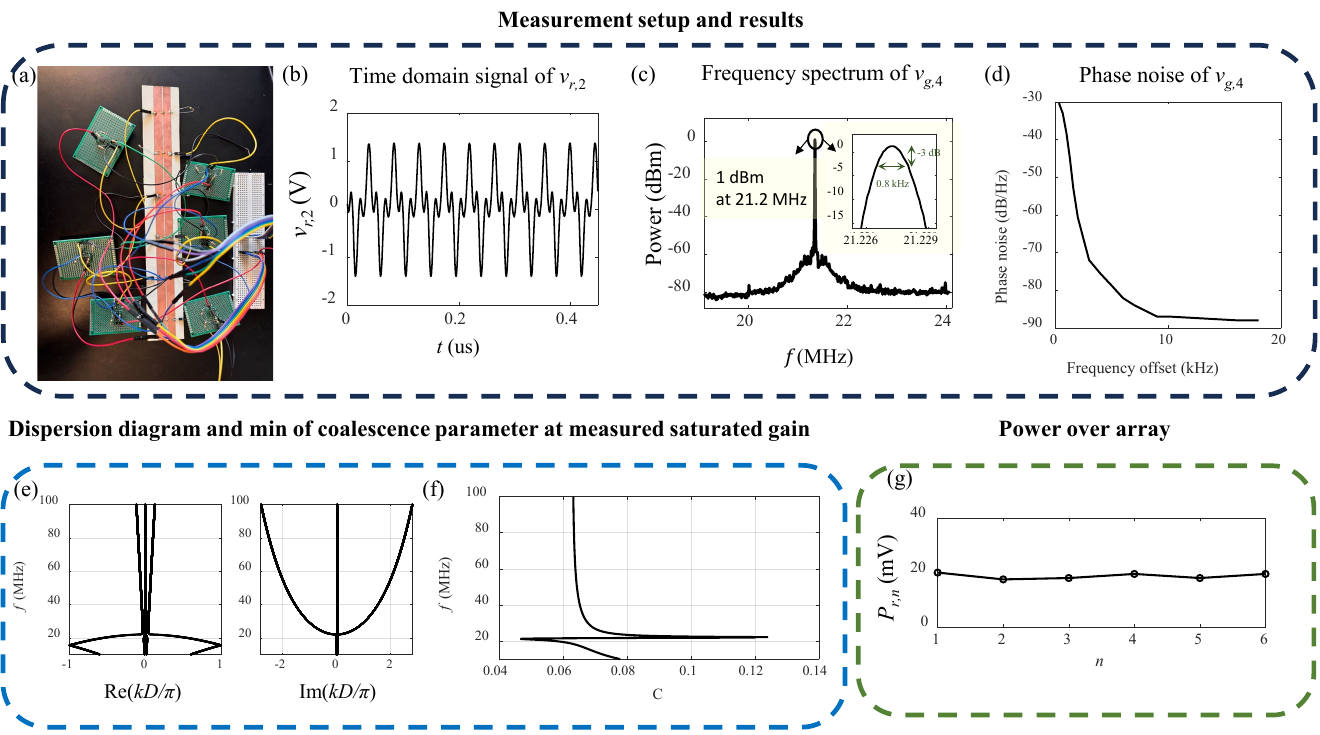}
			\par\end{centering}
		\caption{ (a) Assembled array with a length of $N=6$ (three unit cells). (b) Measured time-domain voltage signal at the fourth array element $Y_{r,2}$ using an oscilloscope. (c) Measured spectrum using a spectrum analyzer (Rigol DSA832E) with the fundamental frequency of oscillation at $21.2\:\text{MHz}$ at the fourth gain element. (d) The phase noise of the power spectrum measured by the spectrum analyzer at frequency offsets from a few Hertz to $20\:\text{kHz}$, with a resolution bandwidth of $100\:\text{Hz}$ and a video bandwidth of $30\:\text{Hz}$ to fully capture the spectrum. (e) Real and imaginary parts of the complex wavenumber $k$ as a function of frequency, showing a second-order EPD at $21.1\:\text{MHz}$ using the saturated gain value $g_{sat}=19\:\text{mS}$. (f) Coalescence parameter confirming at least two eigenvectors coalescence at $21.1\:\text{MHz}$. (g) Measured radiated power along the array.
\label{fig:Measurment}}
\end{figure*}

\subsection{Experimental results}\label{subsec:Measurment}

The experimental board is shown in Fig.~\ref{fig:Measurment}(a). This board implements the structure shown in Fig.~\ref{fig:Measurment_sim}(a), where the nonlinear gain is created with an op amp-based inverter circuit (Texas Instruments, model LMH6702MAX/NOPB), as depicted in Fig.~\ref{fig:Inverter}(a) in Appendix~D. To achieve a small-signal gain of $ g = 21.3 \:\text{mS}$, we used a $47 \:\Omega$ resistor in series with a $1 \:\Omega$ resistor. The additional $1 \:\Omega$ resistor was added to compensate for the loss produced by the dummy $1 \:\Omega$ resistor placed in series with the inverter. This dummy resistor allows us to measure the current entering the inverter system, enabling us to calculate the active admittance in the saturated regime as detailed in Appendix~D.  By setting $g = 21.3\:\text{mS}$, we measure the time domain of $v_{r,2}$ and frequency spectrum signal of $v_{g,4}$, as shown in Fig.~\ref{fig:Measurment}(b)~and~(c), respectively. Based on the measurements discussed in Appendix~D, the negative gain admittance achieved by the inverter circuit is mainly real valued. We measured the oscillation frequency to be $f_\text{osc} = 21.2\:\text{MHz}$, noting a slight shift from the simulation results. We suspect this small difference between the simulated and measured oscillation frequencies can be attributed to different nonlinearity and saturation models.  

\begin{figure}[t]
		\begin{centering}
			\includegraphics[width=3.4in]{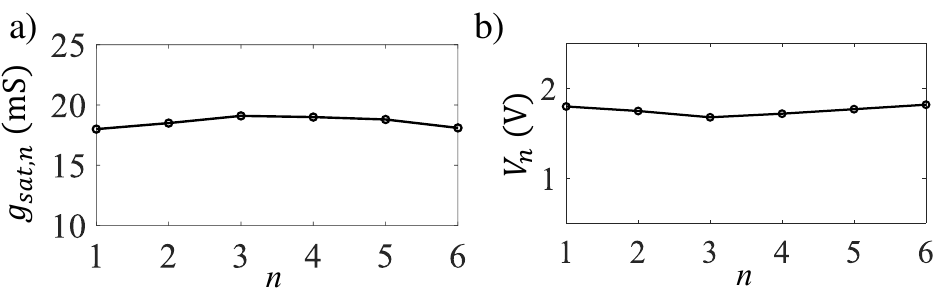}
			\par\end{centering}
		\caption{(a) Measured saturation gain distribution across the array for $N=6$. (b) Voltage amplitudes corresponding to each gain element. They are both more or less constant across the array.
\label{fig:Mesearment_sat}}
\end{figure}

To confirm that the system tends to oscillate at an EPD in the saturation regime, we measured the saturated gain of the active elements. Using the technique discussed in Appendix~D, we measured $g_{sat,4} \approx 19\:\text{mS}$, which is close to the real value found via ADS time-domain simulations. The measured saturation gain $g_{sat}$ across the arrayed active elements and the corresponding voltage amplitudes are shown in Fig.~\ref{fig:Mesearment_sat}(a)~and~(b), respectively. The measured $g_{sat,n}$ shows a uniform gain distribution across the array. 

An essential feature of any oscillator is its ability to produce a near-perfect periodic time-domain signal, which is quantified in terms of phase noise \cite{hajimiri1998general,oveisi2023study}. The phase noise refers to the random fluctuations in the phase of an oscillator's output, and it plays a critical role in determining the stability and quality of the output signal. Minimizing phase noise ensures reliable performance in applications where precise frequency control is necessary \cite{lee2000oscillator}. The measured power spectrum and phase noise up to a $20\:\text{kHz}$ frequency offset are shown in Fig.~\ref{fig:Measurment}(c)~and~(d). Despite the presence of electronic noise (which is significant in the op amp) and thermal noise, the proposed oscillator exhibits stable oscillation with notably small harmonics. The low phase noise of $-89\:\text{dB/Hz}$ at a $10\:\text{kHz}$ offset from the oscillation frequency confirms the oscillator's stability and precision. A key indicator of clean oscillation is the narrow linewidth of the oscillation spectrum in Fig.~\ref{fig:Measurment}(c). We used a resolution bandwidth of $100\:\text{Hz}$ and a video bandwidth of $100\:\text{Hz}$. All the measured frequency spectra of gain over the array remain clean, approximately down to $-81\:\text{dB}$ from the peak value. At the fourth gain element, the linewidth, measured at $-3\:\text{dB}$ from the peak, is around $0.8\:\text{kHz}$, indicating the very narrow linewidth relative to the oscillation frequency.

To confirm the saturated system supports an EPD at the oscillation frequency, we calculate the Bloch mode dispersion diagram using the gain of $g_{sat} = 19\:\text{mS}$ measured in the saturation regime. The real and imaginary parts of the complex Bloch wavenumber $k$ as a function of frequency are shown in Fig.~\ref{fig:Measurment}(e). The plot highlights the presence of a second-order EPD with $k=0$ at $21.1\:\text{MHz}$, which is very close to the oscillation frequency obtained from both the time-domain simulations and the experimental measurements. To further validate the EPD occurrence, we calculate the coalescence parameter $\text{C}$ using the measured saturated gain of $g_{sat} = 19\:\text{mS}$. As shown in Fig.~\ref{fig:Measurment}(f), the coalescence parameter reaches its minimum at $21.1\:\text{MHz}$, verifying the existence of an EPD in the system. 

Figure~\ref{fig:Measurment}(g) shows the measured time-average power delivered to each array element. The power $P_{r,n} = Y_r |V_{r,n}|^2 / 2$ is determined by measuring the voltage $V_{r,n}$ on each real-valued radiation admittance  $Y_r=21.3\:\text{mS}$. Here, $|V_{r,n}|$ represents the phasor magnitude, calculated as the peak value of the measured time-domain signal. For example, the measured time-average power $P_{r,4} = 19.4\:\text{mW}$ delivered to the fourth array element is more or less close to the power generated by the fourth active element $P_{g,4} = g_{sat,4} |V_{g,4}|^2 / 2 = 26.8 \:\text{mW}$.

The plot in Figure~\ref{fig:Measurment}(g) also shows that the power over the array elements is more or less constant across the array. We have used resistors with a $1\%$ tolerance, while the active elements may have an ever larger small-signal gain variation across the array. The uniform power distribution delivered to the admittances $Y_{r,n}$ implies that arrayed antennas working at an EPD will maintain constant radiated power across the array, highlighting the robustness of the design in practical scenarios with elements' variability. These findings show the important result that the array radiates power at a stable, low-noise, oscillation frequency, as discussed in Sec.~\ref{subsec:Simulation_meas}, even when perturbations occur.

\section{Conclusion}
We have demonstrated that the proposed array of active nonlinear elements and loads (representing antennas), achieves a stable oscillation frequency and uniform power distribution over the lossy elements with a non-zero saturated gain across the array. The work in Ref.~\cite{Nikzamir2024Exceptional} showed a degenerate active array structure that saturated gain that decreases when increasing the array length, converging to an EPD. In this work, we overcome that limitation by achieving a periodic array system with non-zero saturated gain across the array, oscillating at a stable frequency, with a narrow linewidth, and with uniform power distribution across the array. The active array consistently operates near an EPD in its steady state, regardless of the initial small-signal gain values. This inherent stability, along with the resilience to variations in gain or loss variations and discrete element failure, underscores the robustness of the new concept. These concepts have been verified via simulations for two array designs, and experimentally for the second design. 

Our findings indicate that the EPD is the desirable operational regime, because of the nonlinear gain dynamics. Indeed, the saturated gain is predominantly uniform across all nonlinear active elements and equal to the one that leads to an EPD in the infinitely long array. The system's tendency to reach a purely real saturated gain, especially in longer structures, further supports the effectiveness of the proposed design in maintaining the exceptional degenerate regime under various conditions. 

By operating at an EPD, the array oscillator not only stabilizes the oscillation frequency with respect to the array length but also provides radiation power that increases with the array length, making it highly suitable for high-power radiation applications requiring a stable frequency with low noise.

\section*{Appendix A: Coefficients for the dispersion relation}
\label{sec:AppendixCoeff}

The coefficients $b$ and $c$ in Eq.~(\ref{eq:FullDispersionRelation}) are

\begin{widetext}
\begin{equation}
  \begin{array}{c}
\text{Re}(b)=(Y_{r}-g)^{2}\left(Z_{o}\sin\theta_{o}-Z_{e}\sin\theta_{e}\right)^{2}/4\\
-2Y_{r}gZ_{e}Z_{o}\sin\theta_{e}\sin\theta_{o}+4\left(1-\cos^{2}\theta_{e}-\cos^{2}\theta_{o}\right)\\
\text{Im}(b)=(g-Y_{r})\left(Z_{e}\sin(2\theta_{e})+Z_{o}\sin(2\theta_{o})\right)\\
\text{Re}(c)=(Y_{r}-g)^{2}\left(Z_{o}\sin\theta_{o}-Z_{e}\sin\theta_{e}\right)^{2}/2-(Y_{r}-g)^{2}\left(Z_{e}\sin\theta_{e}\cos\theta_{o}+Z_{o}\sin\theta_{o}\cos\theta_{e}\right)^{2}+(Y_{r}gZ_{e}Z_{o}\sin\theta_{e}\sin\theta_{o}\\
+4\cos\theta_{e}\cos\theta_{o})^{2}-4Z_{e}Z_{o}\sin\theta_{e}\sin\theta_{o}Y_{r}g-8\cos^{2}(\theta_{o})-8\cos^{2}\theta_{e}+6\\
\text{Im}(c)= (Y_r - g) [ Y_r g Z_o Z_e\left( Z_e \sin(2\theta_{o}) \sin^2\theta_{e} 
        + Z_o \sin(2\theta_{e}) \sin^2\theta_{o} \right) 
        + 2 Z_o \sin(2\theta_{o}) \cos(2\theta_{e}) \\
        + 2 Z_e \sin(2\theta_{e}) \cos(2\theta_{o})],
\end{array}
    \label{eq:CoefficientsDispersionRelation}
\end{equation}
\end{widetext}
\noindent
where $Z_{e}$ and $Z_{o}$ are characteristic impedances and $\theta_e=\beta_e d$ and $\theta_o=\beta_o d$ are the electric lengths of the even and odd modes of the uniform coupled waveguide segments, respectively.

Because of reciprocity, if a complex wavenumber $k$ is a solution to the characteristic equation, $-k$ is also a solution. Therefore, we simplify the characteristic equation as follows: the eigenvalues $\zeta_1$, $1/\zeta_1$, $\zeta_2$ and $1/\zeta_2$ are the four solutions of the characteristic equation that is rewritten as
\begin{equation}
\begin{array}{c}
(\zeta-\zeta_{1})(\zeta-1/\zeta_{1})(\zeta-\zeta_{2})(\zeta-1/\zeta_{2})\\
=(\zeta^{2}-a_1\zeta+1)(\zeta^{2}-a_2\zeta+1)=0,
\end{array}
\label{eq:CharacteristicEquationZetas}
\end{equation}
where 
\begin{equation}
        a_1 = \zeta_{1}+1/\zeta_{1}, \:\:\: a_2 = \zeta_{2}+1/\zeta_{2}.
\end{equation}

Expanding this equation, we write
\begin{equation}
    \zeta^4 - \zeta^3 (a_1 + a_2) + \zeta^2(2+a_1a_2) - \zeta(a_1 + a_2) + 1 = 0,
    \label{eq:a1a2CharacEq}
\end{equation}
which coincides with \ref{eq:FullDispersionRelation}, whose solutions are the four Bloch eigenmodes.

\section*{Appendix B: Resilience to small-signal gain variations}

\begin{figure*}[t]
		\begin{centering}
			\includegraphics[width=7in]{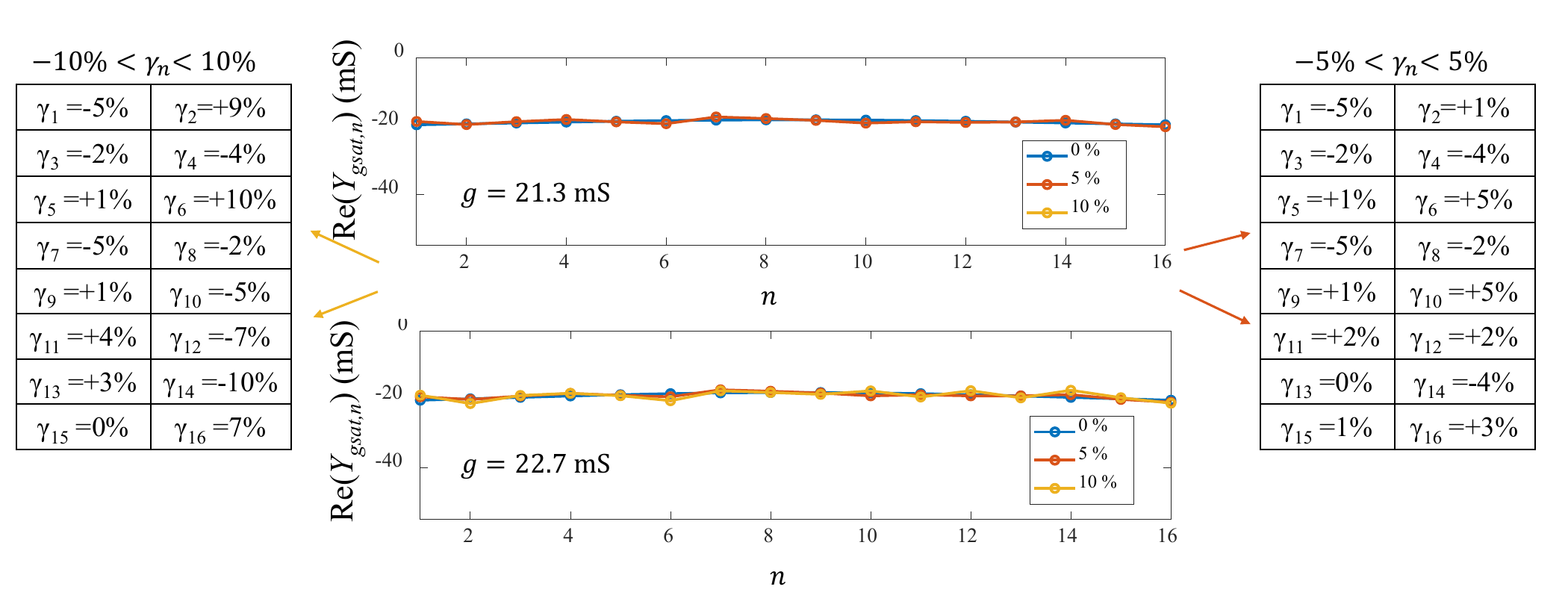}
			\par\end{centering}
		\caption{ Simulated saturated gain of each nonlinear gain element for two small-signal gain values $g=21.3\:\text{mS}$ and $g=22.7\:\text{mS}$. In each case, we apply a random perturbation of the array's small-signal admittances $Y_{g,n}=-g_n$ across the array with relative variation $-5\%<\gamma_n<5\%$ and  $-10\%<\gamma_n<10\%$. The saturated gains are basically the same for all the six cases considered, confirming that the array saturates at and EPD independently of the variations and perturbations of small-signal gain. \label{fig:perturbationSimul}}
	\end{figure*}

To ensure the experiment's resilience to tolerances, we consider scenarios where there are variations in the active elements and investigate via simulations the resilience of the system to such perturbations. We examine two different initial small-signal gains, $g = 21.3\:\text{mS}$ and $22.7\:\text{mS}$. In each case, we assume a large variation of the small-signal gain values from array element to element. Specifically, the small-signal gain varies, for each of the two different scenarios, as $g_{n} = g\times \gamma_{n}$, where $-5\% < \gamma_{n} < 5\%$ or $-10\% < \gamma_{n} < 10\%$. These variations are illustrated in Fig.~\ref{fig:perturbationSimul}. For the cases with maximum perturbation of  $5\%$ or $10\%$, the $\gamma_{n}$ values are shown as an inset. In each scenario, the resulting saturated gain value, obtained via ADS time-domain simulations, is more or less the same and it leads to an EPD at roughly the same frequency. Furthermore, the oscillation frequency in the steady state regime is maintained at $f_\text{osc} = 21.67\:\text{MHz}$ in all six cases.

 \section*{Appendix C: Power Extraction and Oscillation Frequency Stability Varying Array Length}

\begin{figure}[t]
		\begin{centering}
			\includegraphics[width=3.5in]{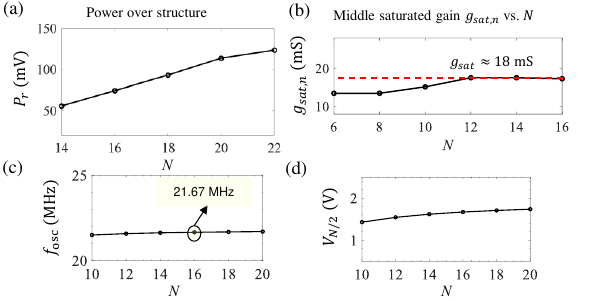}
			\par\end{centering}
		\caption{ (a) Total ``radiated'' power increases with array length, where $N$ is the number of active elements. (b) Calculated real part of the saturated gain in the middle of the array ($n=N/2$) versus array length $N$, which shows a trend where the saturated gain approaches a constant value $g_{sat}\approx18\:\text{mS}$ independently of the array length. (c) The oscillation frequency remains stable with changes in array length, showing less than $1\:\%$ variation when the number of array elements doubled from $N = 10$ to $N = 20$. (d) Oscillation signal's voltage amplitude at the active elements at the center of the array  ($n=N/2$), from $N = 10$ to $N = 20$.\label{fig:power_sims}}
	\end{figure}
 
Here, we demonstrate how the power extraction   (i.e., delivered to the loads $Y_r$)  increases when the array length increases, while maintaining a stable oscillation frequency. To do this, we conducted time-domain simulations using Keysight ADS to determine the total power, and oscillation frequency versus array length. All the results in this Appendix are based on simulations, assuming the small-signal gain is $g=21.3\:\text{mS}$. Additionally, we analyze the saturated gain in the middle of the structure as a function of $N$. Our goal is to confirm that the structure operates at the EPD, independently of the array length.

Figure~\ref{fig:power_sims}(a) illustrates the {\em total} power delivered to all the $N$ elements $Y_r$, showing a clear increase in radiated power as the array lengthens. Moreover, in Fig.~\ref{fig:power_sims}(b) we present the saturated gain in the middle of the structure, i.e., for $n=N/2$, as a function of $N$, revealing a more or less constant  $g_{sat,n} \approx 18\:\text{mS}$, with $n=N/2$, after $N=12$.  Furthermore, Fig.~\ref{fig:power_sims}(c) demonstrates that the oscillation frequency within the array is $f_\text{osc} = 21.67\:\text{MHz}$, confirming that the concept of extracting more power from a longer array with stable oscillation is feasible. The result shows that the oscillation frequency changes by less than $1\:\%$ as the system length increases from $N=10$ to $N=20$. 

Figure~\ref{fig:power_sims}(d) shows the voltage amplitude of the oscillation signal at the middle gain element ($n=N/2$) for array lengths ranging from $N=10$ to $N=20$. For instance, in an array with $N = 14$, the voltage amplitude at the center element is $V_{g,7} = 1.6\:\text{V}$. To further validate the measurement results, the power provided by this active element in the steady state regime is calculated as $P_{g,7} \approx 23\:\text{mW}$, which corresponds to the measured power $P_{r,4}$ delivered on $Y_r$ shown in Fig.~\ref{fig:Measurment}(g). Furthermore, since the system operates in a saturated gain regime with approximately constant gain across the array, the total power produced by all the active elements is expected to follow $P_g \approx N \times P_{g,N/2}$, where $P_g$ is the total power and $P_{g,N/2}$ represents the power delivered by the middle gain element. By analyzing the voltage distribution for different array lengths and the total power, we confirm that the system maintains a uniform operational state due to the EPD.

\section*{Appendix D: Measurement of Negative Conductance in the Saturated Regime}\label{appendix:inverter}

\begin{figure}[t]
		\begin{centering}
			\includegraphics[width=3.5in]{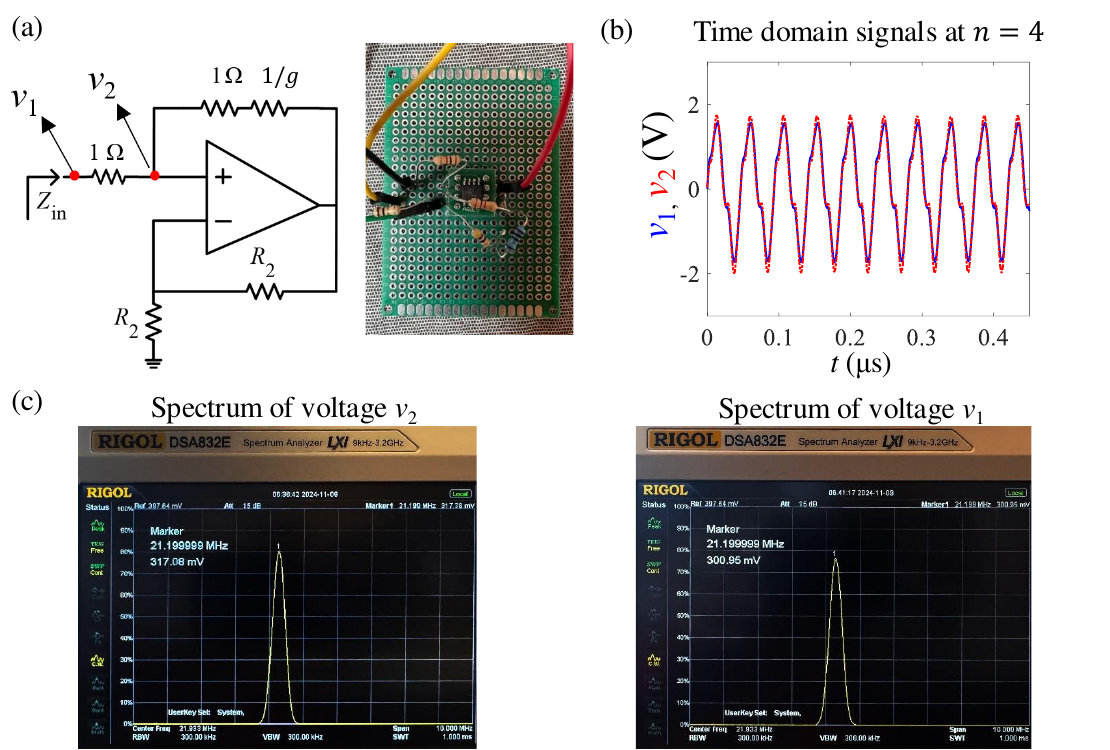}
			\par\end{centering}
		\caption{  (a) Schematic of the circuit using an op amp to provide a negative conductance $Y_g=-g$. (b) Measured time-domain signals $v_1$ and $v_2$ in phase, showing that $Y_{gsat,4}$ is mainly real. (c) The measured frequency spectrum of $v_1$ and $v_2$, showing the oscillation frequency is $21.2\:\text{MHz}$  when $N=6$. \label{fig:Inverter}}
\end{figure}

Several approaches can provide the negative conductances for the proposed array. In the experiment, we have used the circuit shown in Fig.~\ref{fig:Inverter}(a) that utilizes an operational amplifier (op amp) to generate a negative impedance. The converter circuit modifies the impedance such that $Z_{\text{in}} = -(1/g+1) + 1\:\Omega$. In our experiment, we used $1/g=47 \:\Omega$ and $R_2 = 1\:\text{k}\Omega$ to achieve the small-signal gain of $g = 21.3\:\text{mS}$. 

To measure the gain in the saturated regime, we first measured the voltage $v_{\text{in}}=v_1$ at the nonlinear active element in Fig.~\ref{fig:Inverter}(a) in the saturation regime. We also measured the voltage across a dummy $1\:\Omega$ resistor that leads to the calculation of the input current of the active element. Note that we have added another $1\:\Omega$ in the inverter in series to $1/g$ to compensate for the dummy resistance $1\:\Omega$, resulting in $Z_{\text{in}}=-1/g=-1/21.3\:(\text{1/mS})$. 
In Fig.~\ref{fig:Inverter}(b), we show the measured $v_1$ and $v_2$ for the fourth nonlinear gain element with respect to the ground using an oscilloscope. The voltage difference is measured to estimate the current $i_{\text{in}}=(v_1-v_2)/(1 \:\Omega)$; the measured saturated gain is calculated, in phasor terms, as $Y_{gsat}=I_{\text{in}}/V_{\text{in}}$. As expected, the phase shift observed between the voltage $v_2$ and $v_1$ across the $1\:\Omega$ resistor is negligible, as shown in Fig.~\ref{fig:Inverter}(b), leading to an almost real-valued $Y_{gsat}$. To further analyze the voltages $v_1$ and $v_2$, we used a spectrum analyzer to measure their spectrum in Fig.~\ref{fig:Inverter}(c), that confirm that their oscillation frequency is $21.2\:\text{MHz}$. These measurements are done using a resolution bandwidth of $300\:\text{kHz}$ and a video bandwidth of $300\:\text{kHz}$.  Based on the measured voltages, we obtained a saturated gain admittance $g_{sat,4} \approx 19\:\text{mS}$.

\bibliographystyle{IEEEtran}
\bibliography{2ndEPD_TL}
	
\end{document}